\documentclass[
 aip,
 pre,
 amsmath,amssymb,
 reprint
]{revtex4-1}

\usepackage{graphicx}
\usepackage{dcolumn}
\usepackage{bm}

\usepackage[utf8]{inputenc}
\usepackage[T1]{fontenc}
\usepackage{mathptmx}
\usepackage{etoolbox}
\usepackage{xcolor}
\usepackage{comment}
\usepackage{natbib}

\makeatletter
\def\@email#1#2{%
 \endgroup
 \patchcmd{\titleblock@produce}
  {\frontmatter@RRAPformat}
  {\frontmatter@RRAPformat{\produce@RRAP{*#1\href{mailto:#2}{#2}}}\frontmatter@RRAPformat}
  {}{}
}%
\makeatother
\usepackage{hyperref} 
\begin{document}

\preprint{AIP/123-QED}

\title[]{On the consistency of pseudo-potential lattice Boltzmann methods}
\author{Luiz Eduardo Czelusniak}
 \email{luiz.czelusniak@partner.kit.edu}
\affiliation{ 
Lattice Boltzmann Research Group (LBRG), Institute for Applied and Numerical Mathematics (IANM), Karlsruhe Institute of Technology (KIT), Karlsruhe, Baden Württemberg, D-76131, Germany
}%
\author{Tim Niklas Bingert}%
\affiliation{ 
Lattice Boltzmann Research Group (LBRG), Institute of Mechanical Process Engineering (MVM), Karlsruhe Institute of Technology (KIT), Karlsruhe, Baden Württemberg, D-76131, Germany
}%

\author{Mathias J. Krause}
\affiliation{ 
Lattice Boltzmann Research Group (LBRG), Institute for Applied and Numerical Mathematics (IANM), Karlsruhe Institute of Technology (KIT), Karlsruhe, Baden Württemberg, D-76131, Germany
}%
\affiliation{ 
Lattice Boltzmann Research Group (LBRG), Institute of Mechanical Process Engineering (MVM), Karlsruhe Institute of Technology (KIT), Karlsruhe, Baden Württemberg, D-76131, Germany
}%

\author{Stephan Simonis}
\affiliation{ 
Lattice Boltzmann Research Group (LBRG), Institute for Applied and Numerical Mathematics (IANM), Karlsruhe Institute of Technology (KIT), Karlsruhe, Baden Württemberg, D-76131, Germany
}%

\date{\today}

\begin{abstract}
We derive the partial differential equation (PDE) to which the pseudo-potential lattice Boltzmann method (P-LBM) converges under diffusive scaling, providing a rigorous basis for its consistency analysis. By establishing a direct link between the method’s parameters and physical properties—such as phase densities, interface thickness, and surface tension—we develop a framework that enables users to specify fluid properties directly in SI units, eliminating the need for empirical parameter tuning. This allows the simulation of problems with predefined physical properties, ensuring a direct and physically meaningful parametrization.
The proposed approach is implemented in OpenLB, featuring a dedicated unit converter for multiphase problems. To validate the method, we perform benchmark tests—including planar interface, static droplet, Galilean invariance, and two-phase flow between parallel plates—using R134a as the working fluid, with all properties specified in physical units. The results demonstrate that the method achieves second-order convergence to the identified PDE, confirming its numerical consistency. These findings highlight the robustness and practicality of the P-LBM, paving the way for accurate and user-friendly simulations of complex multiphase systems with well-defined physical properties.
\end{abstract}

\maketitle

\section{\label{sec:level1}Introduction}

The lattice Boltzmann method (LBM) has emerged as a promising tool for simulating multiphase systems\cite{gong2013lattice,li2015lattice,cao2022lattice,le2023improved,alsadik2024lattice}, primarily due to its high computational efficiency, particularly in parallel simulations using multicore CPUs\cite{heuveline2009towards,henn2016parallel,marquardt2024novel} and GPUs\cite{cheng2018application,yang2023implementation}. Several multiphase approaches exist within the LBM framework, including the pseudo-potential\cite{shan1993lattice,shan1994simulation}, free-energy\cite{swift1996lattice,guo2002lattice}, phase-field\cite{he1999lattice,lee2005stable,liang2018phase}, and color-gradient methods\cite{rothman1988immiscible,gunstensen1991lattice}. Among these, the pseudo-potential method (P-LBM) is particularly useful for high-performance simulations\cite{qin2022effective,zhao2022pseudopotential} due to its simplicity and efficiency\cite{czelusniak2025fundamental}. The interface naturally emerges from the interparticle forces acting on the fluid\cite{pasieczynski2021fluid}, eliminating the need for interface capturing or tracking methods.

However, many simulations using the P-LBM are conducted using lattice units without establishing a clear correspondence between the simulation and a real physical condition\cite{wang2022unit}. 
\citeauthor{baakeem2021novel}\cite{baakeem2021novel} 
argue that the conversion or mapping of parameters between the macroscopic (or physical) scale and the mesoscopic (or lattice) scale is not a straightforward procedure for the P-LBM. 
Therefore, some authors have focused on the dimensional analysis of this method
\cite{baakeem2021novel,jaramillo2022pseudopotential,wang2022unit,zheng2023improved,oh2024turbulent}. As a result, different strategies for mapping parameters between the macroscopic and mesoscopic scales were proposed.

Despite the progress, some points still make P-LBM challenging to apply to real physical problems. First, the method relies on specifying the properties of the equation of state (EOS), whereas the phase-field method allows for the direct imposition of surface tension and interface thickness, making it more practical to use. Another difficulty mentioned by \citeauthor{czelusniak2025fundamental}\cite{czelusniak2025fundamental} is that the method's parameters, necessary to impose the desired surface tension and ensure thermodynamic consistency, must be found empirically through prior simulations such as the Young--Laplace test.

To address these issues, we revisited the topic of P-LBM consistency.
Through our investigation, we derived the form of the macroscopic equation to which the P-LBM converges under diffusive scaling. This understanding allowed us to directly relate the fluid's physical properties—such as phase densities, interface thickness, and surface tension—to the P-LBM parameters. Consequently, one of the method's major limitations was addressed: Users no longer need to fine-tune specific parameters, but can instead define desired fluid properties directly, greatly simplifying its application.

To implement and validate these developments, we utilized OpenLB\cite{openlb2020} based on version 1.7\cite{olbRelease17}. This software is a powerful tool for parallel lattice Boltzmann simulations that is employed for example in sub-grid particulate flows\cite{bukreev2023simulation}, fully resolved particle flows\cite{hafen2023simulation,marquardt2024novel}, turbulence simulations\cite{siodlaczek2021numerical,simonis2022temporal}, optimization\cite{reinke2022applied,jessberger2022optimization}, sub-grid multiphase flows\cite{bukreev2023consistent} and fully resolved multiphase flows\cite{simonis2024binary}. 
OpenLB employs unit converters, enabling users to specify physical variables directly in physical units that are then transformed into lattice units by the converter. We extended this framework by developing a dedicated converter for multiphase problems. This enhancement allows users to define fluid properties not only directly but also in physical units, making the method significantly more user-friendly and versatile.

To validate our approach, we selected four benchmarks with analytical solutions for diffuse interface problems: planar interface, static droplet, Galilean invariance, and two-phase flow between parallel plates.
All tests are performed in SI units using R134a as a working fluid. Each problem was solved using meshes with varying resolutions, and we demonstrated that the method achieves second-order convergence in all cases, confirming its consistency.

The remainder of this paper is organized as follows. In Section~\ref{sec:PLBM}, we introduce the theoretical foundations of the pseudo-potential lattice Boltzmann method (P-LBM). Section~\ref{sec:Dimension} focuses on the derivation of the partial differential equation (PDE) approximated by the P-LBM under diffusive scaling. In Section~\ref{sec:Parameters}, we present the methodology for directly relating the method parameters to physical properties. Section~\ref{sec:OpenLB} describes the implementation of the proposed framework in OpenLB, including the development of a dedicated unit converter for multiphase problems. Section~\ref{sec:Results} presents the results of the benchmark tests, validating the consistency of the method. Finally, Section~\ref{sec:Conclusion} concludes the paper with a summary of the main findings.


\section{\label{sec:PLBM}Pseudo-potential LBM}

In this work, we apply the lattice Boltzmann equation (LBE) with the single-relaxation-time (SRT or BGK) collision operator\cite{bhatnagar1954model}. To establish the dimensions of each variable, we define the fundamental dimensions of mass ($M$), length ($L$), and time ($T$). We denote the dimension of a variable $\phi$ as $[\phi]$. The LBE is expressed as:
\begin{equation} \label{eq:LBE}
f_i( t + \Delta t, \bm{x} + \bm{c}_i \Delta t ) - f_i( t, \bm{x} )
= - \frac{\Delta t}{\tau} 
( f_i - f_i^{eq} ) + \Delta t F_i',
\end{equation}
where $f_i$ ($[f_i] = ML^{-3}$) are the particle distribution functions associated with particle velocities $\bm{c}_i$, and $f_i^{eq}$ are the local equilibrium distribution functions (EDF). The BGK collision operator is governed by the relaxation time $\tau$ ($[\tau] = T$).

The variables t and $\bm{x}$ are the time and space coordinates, respectively. The velocity set used is the regular two-dimensional nine velocities scheme (D2Q9):
\begin{equation} 
\label{eq:VelocitySet}
\bm{c}_i =
\begin{cases} 
      (0,0), ~~~~~~~~~~~~~~~~~~~~~~~~~~~~~~~~~~~~~~~~~~ i = 0,  \\
      (c,0), (0,c), (-c,0), (0,-c), ~~~~~~ i = 1,...,4, \\
      (c,c), (-c,c), (-c,-c), (c,-c), ~ i = 5,...,8, \\
   \end{cases}
\end{equation}	
where $c$ is the lattice speed defined as $c=\Delta x/\Delta t$. The parameters $\Delta x$ and $\Delta t$ are the space and time steps.

The specific form of $f_i^{eq}$ ($[f_i^{eq}] = ML^{-3}$) varies depending on the particular method employed. In this work, we use the form:
\begin{equation} \label{eq:Single_Phase_EDF}
    f_i^{eq} = w_i \rho \left( 
    1 + \frac{c_{i\alpha}}{c_s^2}u_{\alpha}
    + \frac{c_{i\alpha}c_{i\beta}-c_s^2\delta_{\alpha\beta}}{2c_s^4}
    u_{\alpha}u_{\beta},
    \right),
\end{equation}
where $w_i$ (dimensionless) are the weights associated with each velocity $\bm{c}_i$. For the D2Q9 scheme, these weights are given as $w_0 = 4/9$, $w_{1,2,3,4} = 1/9$, and $w_{5,6,7,8} = 1/36$. The parameter $c_s$, referred to as the lattice sound speed, is defined as $c / \sqrt{3}$. The variable 
$\rho$ represents the density, and $u_{\alpha}$ denotes the fluid velocity.

The term $F_i'$ on the right-hand side of Eq.~(\ref{eq:LBE}) represents the forcing scheme. This term incorporates the effects of an external volumetric force field $F_{\alpha}$ ($[F_{\alpha}] = ML^{-2}T^{-2}$) into the macroscopic conservation equations. \citeauthor{shan1993lattice} \cite{shan1993lattice} introduced an interaction force based on nearest-neighbor interactions (see \citeauthor{shan2008pressure} \cite{shan2008pressure} for the definition of nearest-neighbor interactions):
\begin{equation} \label{eq:Shan_Chen_Force}
    F_\alpha^{SC} = - G \psi  ( \mathbf{x} )
\sum_{i=1}^{8} \omega ( | \mathbf{c}_i |^2 ) \psi ( \mathbf{x} + \mathbf{c}_i\Delta t ) \frac{c_{i\alpha}\Delta t}{(\Delta t)^2},
\end{equation}
where $\psi$ is a density-dependent interaction potential, and $G$ is a parameter that controls the strength of the interaction. The parameters $\omega(|\bm{c}_i|^2)$ (dimensionless) are weights, typically set as $\omega(1) = 1/3$ and $\omega(2) = 1/12$. The definition of the interaction potential used in this study follows the approach proposed by \citeauthor{yuan2006equations} \cite{yuan2006equations}, which enables the incorporation of arbitrary equations of state $p_{EOS}$ into the system:
\begin{equation} \label{eq:Interaction_Potential}
    \psi(\rho) = \sqrt{ \frac{2\left( p_{EOS} - \rho c_s^2 \right)}{Gc^2} }
    = \sqrt{ \frac{2\left( p_{EOS}\frac{(\Delta t)^2}{(\Delta x)^2}  - \frac{\rho}{3} \right)}{G} }.
\end{equation}
When this technique is used, the parameter $G$ no longer controls the interaction strength. Since authors in the literature commonly adopt $p_{EOS} < \rho c_s^2$, we can simply set $G = -1$. If $G$ is defined as a dimensionless quantity, then $[\psi] = [\sqrt{\rho}] = M^{1/2}L^{-3/2}$.

When applying the original force proposed by \citeauthor{shan1993lattice}\cite{shan1993lattice}, the pseudo-potential method exhibits certain issues related to thermodynamic consistency and the inability to adjust surface tension. To address these limitations, several approaches were developed\cite{li2012forcing,li2013achieving,li2013lattice,lycett2015improved}. In this work, we add the terms derived by \citeauthor{czelusniak2020force} \cite{czelusniak2020force} into the interaction force:
\begin{equation} \label{eq:interactionForce}
\begin{aligned}
F_{\alpha}^{\text{int}} = F_{\alpha}^{SC} &- \frac{G}{12}\frac{(\Delta x)^4}{(\Delta t)^2}
\left( \frac{3}{2} \epsilon + 2\kappa - 2 \right)
M_{\beta}^{(1)} M_{\alpha\beta}^{(2)} \\
&+ \frac{G}{6}\frac{(\Delta x)^4}{(\Delta t)^2} (\kappa-1)
M_{\alpha}^{(1)} M_{\beta\beta}^{(2)},
\end{aligned}
\end{equation}
where $\epsilon$ and $\kappa$ are adjustable parameters to control phase densities and surface tension. Also, $M_{\alpha}^{(1)}$ and $M_{\alpha\beta}^{(2)}$ are defined as:
\begin{subequations}
\begin{equation}
\begin{aligned}
M_{\alpha}^{(1)} &= \sum_i w_i \psi(\mathbf{x}+\mathbf{c}_i\Delta t)
\frac{ c_{i\alpha} }{ c_s^2 \Delta t} \\
&= \partial_{\alpha}\psi 
+ O\left( (\Delta x)^2 \right),
\end{aligned}
\end{equation}
\begin{equation}
\begin{aligned}
M_{\alpha\beta}^{(2)} &= \sum_i w_i \psi(\mathbf{x}+\mathbf{c}_i\Delta t) 
\frac{ c_{i\alpha}c_{i\beta} - c_s^2\delta_{\alpha\beta} }{ c_s^4(\Delta t)^2 } \\
&= \partial_{\alpha}\partial_{\beta} \psi
+ O\left( (\Delta x)^2 \right).
\end{aligned}
\end{equation}
\end{subequations}

The interaction force can be implemented using various forcing schemes. In this study, we adopt the forcing scheme proposed by \citeauthor{guo2002lattice} \cite{guo2002lattice}:
\begin{equation} \label{eq:Guo_Forcing_Scheme}
F_i' = w_i \left( 1 - \frac{\Delta t}{2\tau} \right) \left( \frac{c_{i\alpha}}{c_s^2}F_{\alpha} + \frac{c_{i\alpha}c_{i\beta}-c_s^2\delta_{\alpha\beta}}{c_s^4}F_{\alpha}u_{\beta} \right).
\end{equation}
Here, the total force $F_{\alpha}$ is the combination of the interaction force $F_{\alpha}^{\text{int}}$ and other external forces, such as gravity.

The relation between the particle distribution functions $f_i$ and the macroscopic fluid velocity $u_{\alpha}$ depends on the chosen forcing scheme. For the scheme adopted in this study, the density and velocity fields are defined as follows:
\begin{subequations}
\begin{equation} \label{eq:Density}
\rho = \sum_i f_i,
\end{equation}	
\begin{equation} \label{eq:Momentum}
\rho u_{\alpha} = \sum_i f_i c_{i\alpha} + \frac{F_{\alpha}\Delta t}{2}.
\end{equation}	
\end{subequations}



\section{\label{sec:Dimension} P-LBM partial differential equation}

When analyzing the pseudo-potential method's numerical scheme using the third-order analysis proposed by \citeauthor{lycett2015improved}\cite{lycett2015improved}, the following mass and momentum conservation equations are obtained:
\begin{subequations}
\begin{equation} 
\label{eq:MassConservation}
\partial_t \rho + \partial_{\alpha} (\rho u_{\alpha}) = 0,
\end{equation}	
\begin{equation} 
\label{eq:MomentumConservation}
\partial_t (\rho u_{\alpha}) + \partial_{\beta} (\rho u_{\alpha} u_{\beta}) =
- \partial_{\beta} p_{\alpha\beta}
+ \partial_{\beta} \sigma_{\alpha\beta}' + F_{g,\alpha}.
\end{equation}	
\end{subequations}
Here, the viscous stress tensor $\sigma_{\alpha\beta}'$ is expressed as:

\begin{equation} \label{eq:Stress_Tensor}
    \sigma_{\alpha\beta}' = 
    \rho \nu \left( \partial_{\beta} u_{\alpha} + \partial_{\alpha} u_{\beta} \right).
\end{equation}

The kinematic viscosity $\nu$, which appear in Eq.~(\ref{eq:Stress_Tensor}), is related to the relaxation time by $\nu=c_s^2(\tau-0.5\Delta t)$.

The effect of the interaction force was incorporated into the pressure tensor $p_{\alpha\beta}$, leaving only the gravitational force $F_{g,\alpha}$ in Eq.~(\ref{eq:MomentumConservation}). The pressure tensor is expressed as:
\begin{equation} \label{eq:New_Pressure_Tensor}
    \begin{aligned}
        p_{\alpha\beta} = \bigg(
        p_{EOS} &- \frac{G}{8}\frac{(\Delta x)^4}{(\Delta t)^2}\epsilon 
        ( \partial_{\gamma} \psi )( \partial_{\gamma} \psi )
        \\&
        + \frac{G}{12}\frac{(\Delta x)^4}{(\Delta t)^2} (3-2\kappa) \psi \partial_{\gamma} \partial_{\gamma} \psi 
        \bigg) \delta_{\alpha\beta} \\ &   
        + \frac{G}{6}\frac{(\Delta x)^4}{(\Delta t)^2} \kappa \psi \partial_{\alpha} \partial_{\beta} \psi.
    \end{aligned}
\end{equation}
In the finite-difference literature, these PDEs would be classified as modified equations\cite{warming1974modified}. It is the differential equation obtained by replacing the numerical approximations in the discrete equation with their Taylor series expansions. The terms retained in the equations above are of the order $(\Delta x)^2$ and $(\Delta t)^2$. A rigorous consistency analysis of the method would require determining the truncation errors, which could be done following the procedure developed by \citeauthor{simonis2022limit}\cite{simonis2022limit}. However, we defer this more in-depth theoretical analysis to future work and proceed with our analysis based on modified equations.

The PDE approximated by the P-LBM is grid-dependent, as it relies on $\Delta x$ and $\Delta t$. This dependency manifests in the pressure tensor (Eq.~\ref{eq:New_Pressure_Tensor}) and the interaction potential $\psi$ (Eq.~\ref{eq:Interaction_Potential}). This behavior introduces challenges when discussing the consistency of the P-LBM. By definition, consistency requires that a numerical method approximates a specific PDE. However, in the case of the P-LBM, the PDE varies with each grid resolution. 

We proceed to analyze the behavior of the PDE under a diffusive scaling, $\Delta t \propto (\Delta x)^2$. For simplicity, we assume $\Delta t = \sqrt{r} (\Delta x)^2$, where $r$ is a constant. Based on this assumption, the following result is obtained:
\begin{equation}
    \lim_{ \Delta x \to 0 } \psi = \lim_{ \Delta x \to 0 } \sqrt{ \frac{2\left( p_{EOS}(\Delta x)^2r  - \frac{\rho}{3} \right)}{G} }
    = \sqrt{\frac{2}{3}}\sqrt{\rho}.
\end{equation}
We assume $G = -1$. Consequently, Eq.~(\ref{eq:New_Pressure_Tensor}) can be reformulated to express the pressure tensor of the PDE solved by the P-LBM:
\begin{equation} \label{eq:newTensor}
    \begin{aligned}
        p_{\alpha\beta} = \bigg(
        p_{EOS} &+ \frac{1}{12r} \epsilon
        ( \partial_{\gamma} \sqrt{\rho} )( \partial_{\gamma} \sqrt{\rho} )
        \\&
        - \frac{1}{18r} (3-2\kappa) \sqrt{\rho} \partial_{\gamma} \partial_{\gamma} \sqrt{\rho} 
        \bigg) \delta_{\alpha\beta} \\ &   
        - \frac{1}{9r} \kappa \sqrt{\rho} \partial_{\alpha} \partial_{\beta} \sqrt{\rho}.
    \end{aligned}
\end{equation}
The presence of the constant $r$ implies that the final pressure tensor remains dependent on the choice of the grid. However, it is possible to compare the results of two simulations under a diffusive scaling refinement, as the pressure tensor remains unchanged. This allows us to determine the convergence order of the method. 

The main advantage of expressing the pressure of the P-LBM PDE in the form of Eq.~(\ref{eq:newTensor}), lies in the ability to directly relate the parameters of the pseudopotential method to the physical properties we aim to impose in the simulation (phase densities, interface thickness and surface tension). This will be explored in Section~\ref{sec:Parameters}.



\section{\label{sec:Parameters} Computing P-LBM parameters}

Obtaining the limiting PDE of the P-LBM is more than a mere theoretical curiosity. In this section, we demonstrate that through the derived PDE, it is possible to relate the multiphase physical properties (phase densities, interface thickness, and surface tension) to the method's parameters (equation of state parameters, $\epsilon$ and $\kappa$). This approach addresses one of the major challenges of the P-LBM: the empirical adjustment of the method parameters to achieve the desired physical properties.

The mechanical stability condition\cite{shan2008pressure,li2012forcing} for the P-LBM is given by:
\begin{equation}
    \int_{\rho_v}^{\rho_l}(p_{EOS}-p_0)\frac{\dot{\psi}}{\psi^{1+\epsilon}}d\rho \propto \int_{\rho_v}^{\rho_l}(p_{EOS}-p_0)\frac{d\rho}{\sqrt{\rho}^{2+\epsilon}} = 0.
\end{equation}
This relation becomes equivalent to the Maxwell equal area rule when $\epsilon = 2$. Throughout the remainder of this work, we assume $\epsilon = 2$ for the numerical tests.

For this work, we propose the following novel equation of state:
\begin{equation}
\begin{aligned}
\label{eq:EOS}
p_{EOS} = \frac{p_c}{\rho_c^3} \Big[
&2\rho^3 + (\rho_v-2\rho_l)\rho^2 - 3\sqrt{\rho_v}\rho^{2.5} \\
&+ 2\sqrt{\rho_v}\rho_l\rho^{1.5} +\sqrt{\rho_v}\rho_l^2\rho^{0.5}
\Big],
\end{aligned}
\end{equation}
where $\rho_v$, $\rho_l$, and $\rho_c$ represent the vapor, liquid, and critical densities, respectively. The critical pressure is denoted by $p_c$. This equation is designed to satisfy the Maxwell construction, with $\rho_v$ and $\rho_l$ as the equilibrium densities. This EOS was proposed because it provides a useful and easy way to link its parameters with macroscopic physical properties.

Next, we want to impose a certain interface thickness $\xi$. The interface is defined as the region where $\rho_v+0.33(\rho_l-\rho_v)<\rho<\rho_v+0.67(\rho_l-\rho_v)$. This can be done by setting $p_c/\rho_c^3$ according to the following formula:
\begin{equation} \label{eq:thickness}
\frac{p_c}{\rho_c^3} = \frac{\left( f_{\xi}(\rho_{\text{I};1})-f_{\xi}(\rho_{\text{I};2}) \right)^2}{6r\xi^2},
\end{equation}
where $\rho_{\text{I};1}=\rho_v+0.33(\rho_l-\rho_v)$ and $\rho_{\text{I};2}=\rho_v+0.67(\rho_l-\rho_v)$. The function $f_{\xi}(\rho)$ is defined as: 
\begin{equation}
\begin{aligned}
f_{\xi}(\rho) &= \frac{1}{\rho_l-\rho_v}\text{ln}
\left( \sqrt{\rho}-\sqrt{\rho_v} \right) \\
&- \frac{1}{2\sqrt{\rho_l}(\sqrt{\rho_l}-\sqrt{\rho_v})}\text{ln}
\left( \sqrt{\rho_l}-\sqrt{\rho} \right) \\
&- \frac{1}{2\sqrt{\rho_l}(\sqrt{\rho_v}+\sqrt{\rho_l})}\text{ln}
\left( \sqrt{\rho} + \sqrt{\rho_l} \right).
\end{aligned}
\end{equation}
The theoretical derivation of this expression for the interface thickness is provided in Appendix~\ref{sec:Interface}.

Finally, the surface tension $\gamma$ is imposed by computing the parameter $\kappa$ as:
\begin{equation} \label{eq:surface_tension}
\kappa = \sqrt{ \frac{27\rho_c^3}{2rp_c} } \frac{\gamma}{f_{\gamma}(\rho_v) -f_{\gamma}(\rho_l)},
\end{equation}
where:
\begin{equation}
f_{\gamma}(\rho) = \frac{\rho^2}{4} - \frac{\rho_2}{2}\rho 
- \frac{\sqrt{\rho_1}}{3}\rho^{1.5} + \sqrt{\rho_1}\rho_2 \rho^{0.5}.
\end{equation}
The theoretical derivation of this expression for surface tension is provided in Appendix~\ref{sec:Surface}. 

The proposed approach does not alter the numerical scheme of the pseudo-potential method, but rather provides a way to link the simulation parameters with the desired physical properties. As a result, the positive features of the method, such as its computational efficiency, are preserved. The pseudo-potential method in the presented formulation is highly efficient, as it uses only one distribution function and a single LBE. Moreover, the interaction force depends only on nearest-neighboring nodes, avoiding costly communication operations between distant lattices \cite{czelusniak2025fundamental}. Thus, one of the appealing aspects of the proposed approach is that it does not add any complexity to the method.

Next, we present details of our OpenLB implementation.



\section{\label{sec:OpenLB} Implementations in OpenLB}

OpenLB\cite{openlb2020} is a robust simulation software based on LBM. In this work, we detail how our implementation operates using physical units.
To make a clear distinction between unit systems we define a generic variable $X_{phs}$ as given in physical units and $X_{lat}$ in lattice units.

First, the user specifies the desired fluid properties in physical units, such as the densities $\rho_{v;phs}$ and $\rho_{l;phs}$, the viscosities 
$\nu_{v;phs}$ and 
$\nu_{l;phs}$, and the surface tension 
$\gamma_{phs}$. Additionally, the user defines the geometry of the domain in physical units. For instance, the user may define a characteristic length of the domain $L_{phs}$. 

To enable unit conversion, the multiphase unit converter requires three key inputs in lattice units. These inputs are: the characteristic length $L_{lat}$, the kinematic viscosity of the liquid phase $\nu_{l;lat}$ and the liquid density 
$\rho_{l;lat}$.
Based on these inputs, the conversion factors for length $C_L$, time $C_T$, and mass $C_M$ from lattice to physical units are calculated:
\begin{equation} \label{eq:unit}
C_L = \frac{L_{phs}}{L_{lat}}, \quad C_T = (C_L)^2\frac{\nu_{l;lat}}{\nu_{l;phs}}, \quad C_M = (C_L)^3 \frac{\rho_{l;phs}}{\rho_{l;lat}}.
\end{equation}
For a pure hydrodynamic problem (without heat transfer or chemical reactions), these conversion factors are sufficient to convert any physical property.  
For example, the conversion factors of velocity $C_U$, volumetric force $C_{VF}$ and surface tension $C_{\gamma}$ are obtained as:
\begin{equation} \label{eq:unit2}
C_U = \frac{C_L}{C_T}; \quad C_{VF} = \frac{C_M}{C_L^2C_T^2}; \quad 
C_{\gamma} = \frac{C_M}{C_T^2}.
\end{equation}
It is worth noting that the conversion factor transforms a variable from lattice units to physical units. For example, if $U_{phs}$ is the velocity in physical units and $U_{lat}$ is the velocity in lattice units, they are related by $U_{phs}=C_U U_{lat}$.

\begin{figure}[ht!]
     \centering
    \includegraphics[width=0.5\textwidth]{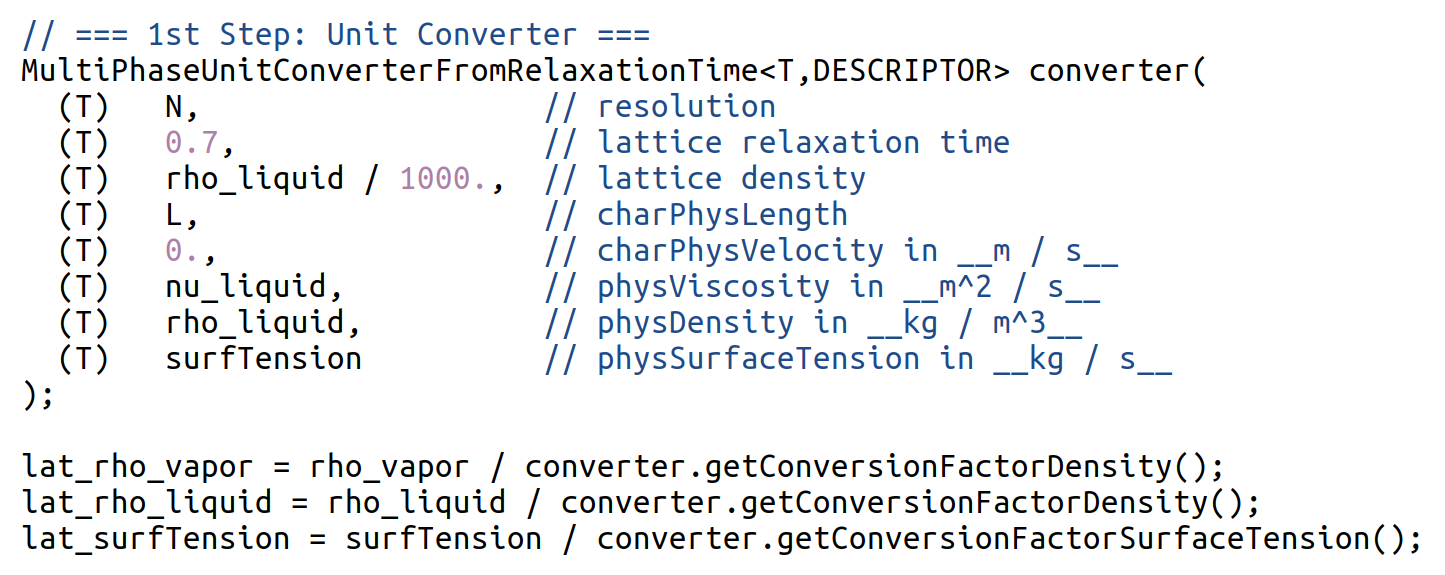}
    \caption{Example of the unit converter implemented in OpenLB, showing the conversion of physical properties into lattice units. This converter allows users to specify fluid properties directly in physical units, simplifying the simulation setup.}
\label{fig:UnitConverter}
\end{figure}

After obtaining all properties in lattice units, such as the interface thickness $\xi$ and surface tension $\gamma$, Eqs.~(\ref{eq:thickness}) and (\ref{eq:surface_tension}) are used to compute $p_c/\rho_c^3$ and $\kappa$. The knowledge of $r$ is not required, as $r = 1$ in lattice units.  
We illustrate our unit converter by presenting a snapshot of our OpenLB code in Fig.~\ref{fig:UnitConverter}. The converter also accepts the characteristic physical velocity as an input if the user wishes to calculate and print the Reynolds number.

The codes used in this work will be available in the next release of OpenLB (1.8), which can be downloaded from the website: \url{https://www.openlb.net/download/}. Authors can also request the codes by email or through the OpenLB forum: \url{https://www.openlb.net/forum/}.



\section{\label{sec:Results}Results}

In this section, we conduct several benchmark tests and analyze the order of convergence of the numerical method with respect to the analytical solutions. All simulations were performed using R134a as the working fluid. Table~\ref{tab:Fluid_Properties} summarizes the fluid properties. 

\begin{table}[ht!]
\centering\small
\renewcommand{\arraystretch}{1.2}
\begin{tabular}{ c | c | c }
\hline
Property
& 
Vapor phase
& 
Liquid phase
\\
\hline
\begin{tabular}{@{}c@{}} Density \\
$\rho$ [$\mathrm{kg/m^3}$] \end{tabular} 
& 
37.5
& 
1187 \\
\hline
\begin{tabular}{@{}c@{}} Kinematic viscosity \\
$\nu$ [$\mathrm{\mu m^2/s}$] \end{tabular} 
&
$0.339$
& 
$0.158$ \\
\hline
\begin{tabular}{@{}c@{}} Surface tension \\
$\gamma$ [$\mathrm{mN/m}$] \end{tabular}  
&
\multicolumn{2}{c}{ $7.58$ }
\\
\hline
\end{tabular}
\caption{Properties of saturated R134a\cite{chae1990surface,oliveira1993viscosity,oliveira1999viscosity} at $T=30^oC$.}
\label{tab:Fluid_Properties}
\end{table}

Unless otherwise stated, we set $\tau_{l;lat}=0.56$ and $\rho_{l;lat}=1.187$ in lattice units. The choice of a small relaxation time was necessary due to the low viscosity of the fluid. Other parameters are computed using properties in physical units and conversion factors -- Eqs.~(\ref{eq:unit}) and (\ref{eq:unit2}). The relaxation time applied in the LBE is a linear combination between the liquid and vapor relaxation times, $\tau_l$ and $\tau_v$:
\begin{equation}
\tau=\tau_v+\frac{\rho-\rho_v}{\rho_l-\rho_v}(\tau_l-\tau_v).
\end{equation}

To analyze the convergence of the method, we define the numerical solution \( s_{\text{num}} \), which is given by a matrix resulting from a simulation and may represent a field of densities, velocities, or other quantities. Based on this, we define the analytical solution as \( s_{\text{ana}} \), which represents a matrix containing the values of the analytical solutions at the same points as \( s_{\text{num}} \). Finally, the error matrix \( e = s_{\text{num}} - s_{\text{ana}} \) is defined. The relative norms are then computed as follows:
\begin{equation} \label{eq:relative_norms}
\begin{aligned}
L_1 &= \frac{\sum|e_{i,j}|}{\sum|s_{\text{ana};i,j}|}, \\
L_2 &= \frac{\sqrt{\sum(e_{i,j})^2}}{\sqrt{\sum(s_{\text{ana};i,j})^2}}, \\
L_{\infty} &= \frac{\text{max}|e_{i,j}|}{\text{max}|s_{\text{ana};i,j}|},
\end{aligned}
\end{equation}
where $i$ and $j$ represent the indexes of the elements of $e$.

In cases where the analytical solution is zero everywhere, such as for spurious currents, we instead define the norms solely in terms of the errors:
\begin{equation} \label{eq:norms}
\begin{aligned}
L_1 &= \sum|e_{i,j}|, \\
L_2 &= \sqrt{\sum(e_{i,j})^2}, \\
L_{\infty} &= \text{max}|e_{i,j}|.
\end{aligned}
\end{equation}

Before describing the tests performed, we provide some comments regarding the scales of the selected problems. We chose problems with characteristic lengths on the order of $10~\mathrm{\mu m}$. Although problems can be simulated on larger scales, the time conversion factor $C_T$ increases proportionally to $C_L^2$ \eqref{eq:unit}, which in turn leads to a reduction in the surface tension conversion factor $C_{\gamma}$ \eqref{eq:unit2}. To recover the correct physical surface tension, it would be necessary to simulate larger surface tension values in lattice units. This can be achieved by increasing the parameter $\kappa$ \eqref{eq:interactionForce} or by reducing the relaxation time to decrease $C_T$.

However, since we are using the BGK collision operator, there is a lower bound on how much we can reduce $\tau$, limiting our ability to adjust $C_T$ freely. We explore the variation of $\kappa$ in the droplet test. For this reason, we opted to work with problems at a relatively small physical scale. We intend to explore problems at different scales in future work.


\subsection{\label{sec:Planar}Planar interface}

\begin{figure}[ht!]
     \centering
     \includegraphics[width=0.48\textwidth]{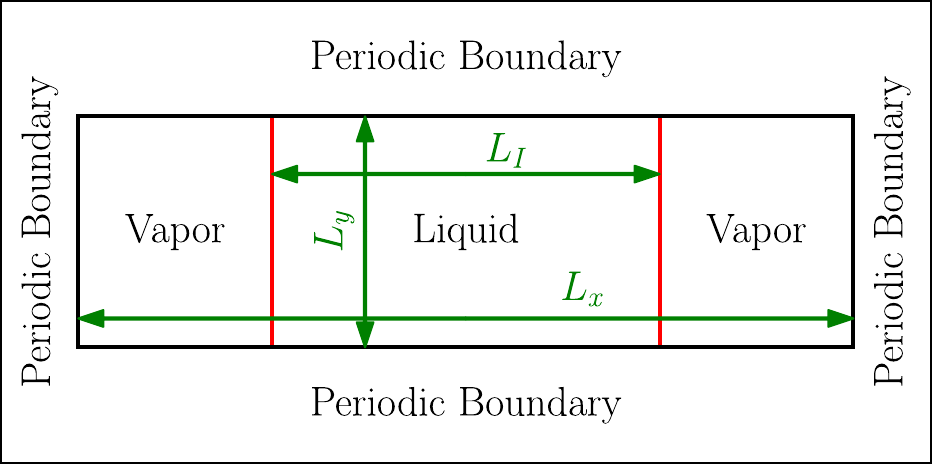}
    \caption{Schematic diagram of the planar interface test configuration. The geometry consists of a two-dimensional domain $(L_x \times L_y)$, where the liquid phase occupies a central region of length $L_I$, surrounded by vapor. Periodic conditions are applied to all boundaries.}
\label{fig:PlanarInterfaceSchematic}
\end{figure}

The first benchmark is the planar interface test problem. Although simple, this test is used here to demonstrate the validity of the equations presented in Section~\ref{sec:Parameters}, particularly regarding the ability to directly prescribe the interface thickness within the method.
A schematic of the test geometry is shown in Fig.~(\ref{fig:PlanarInterfaceSchematic}), where $L_x=10~\mathrm{\mu m}$, $L_y=0.5~\mathrm{\mu m}$ and $L_{I}=5~\mathrm{\mu m}$. Periodic boundary conditions are applied and the simulation is carried out until equilibrium.

The simulation is carried out using five different grid resolutions, and all the necessary information to reproduce the test is shown in Table~\ref{tab:Planar_Interface_Parameters}. 

\begin{table}[ht!]
\centering\small
\renewcommand{\arraystretch}{1.2}
\begin{tabular}{ c | c | c | c | c | c }
\hline
- & Grid 1 & Grid 2 & Grid 3 &
Grid 4 & Grid 5 \\
\hline
$N_x$ & 100 & 200 & 300 & 400 & 500 \\
\hline
$N_y$ & 5 & 10 & 15 & 20 & 25 \\
\hline
$\Delta x$ [nm] & 100 & 50 & 33.3 & 25 & 20 \\
\hline
$\Delta t$ [ps] & 
1266 & 316.4 & 
140.7 & 79.11 & 50.63 \\
\hline
$\xi$ [m] & $1.5\Delta x$ & $3.0\Delta x$ & 
$4.5\Delta x$ & 
$6.0\Delta x$ & $7.5\Delta x$ \\
\hline
$C_U$ [m/s per l.u.] & 79 & 158 & 237 & 316 & 395 \\
\hline
$C_{\gamma}$ [N/m per l.u.] & 0.62 & 1.25 & 1.87 & 2.50 & 3.12 \\
\hline
\end{tabular}
\caption{Discretization parameters of planar interface test case. $N_x$ and $N_y$ are the number of grid nodes used to discretize the domain. $\Delta x$ (in nanometers) is defined as $L_x/N_x$. $\Delta t$ (in picoseconds) is the time step and $\xi$ is the interface thickness. $C_U$ and $C_{\gamma}$ are the conversion factors from lattice units (l.u.) to physical units for velocity and surface tension -- Eq.~(\ref{eq:unit2}). }
\label{tab:Planar_Interface_Parameters}
\end{table}

\begin{figure}[ht!]
     \centering
     \includegraphics[width=0.48\textwidth]{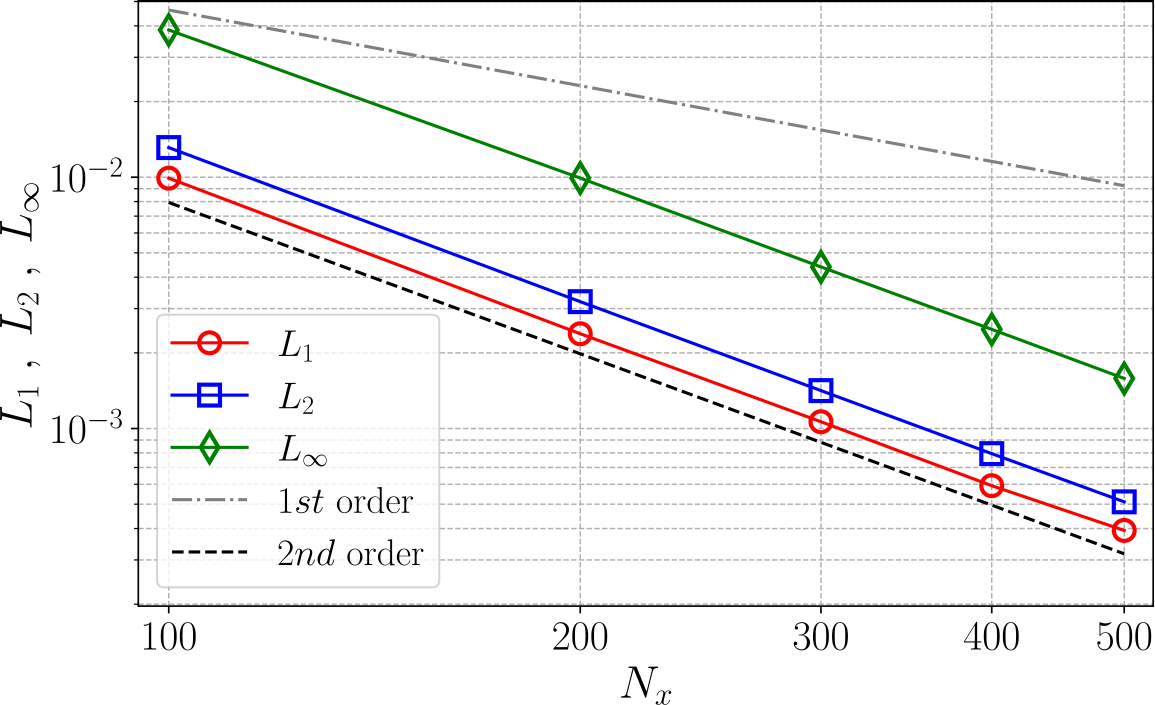}
    \caption{Comparison of the $L_1$, $L_2$, and $L_{\infty}$ norm errors between the numerical simulation and the analytical solution for the planar interface test. The norms are defined in Eq.~(\ref{eq:relative_norms}). Norms are plotted as a function of number of grid points in $x$-direction $N_x$.}
\label{fig:PlanarInterfaceResults}
\end{figure}

The simulation results are presented in Fig.~\ref{fig:PlanarInterfaceResults}. The entire density profile was compared with the analytical solution Eq.~(\ref{eq:analytical_interface}) in Appendix~\ref{sec:Interface}. The error norms converge with second order of accuracy.

\begin{figure}[ht!]
     \centering
     \includegraphics[width=0.48\textwidth]{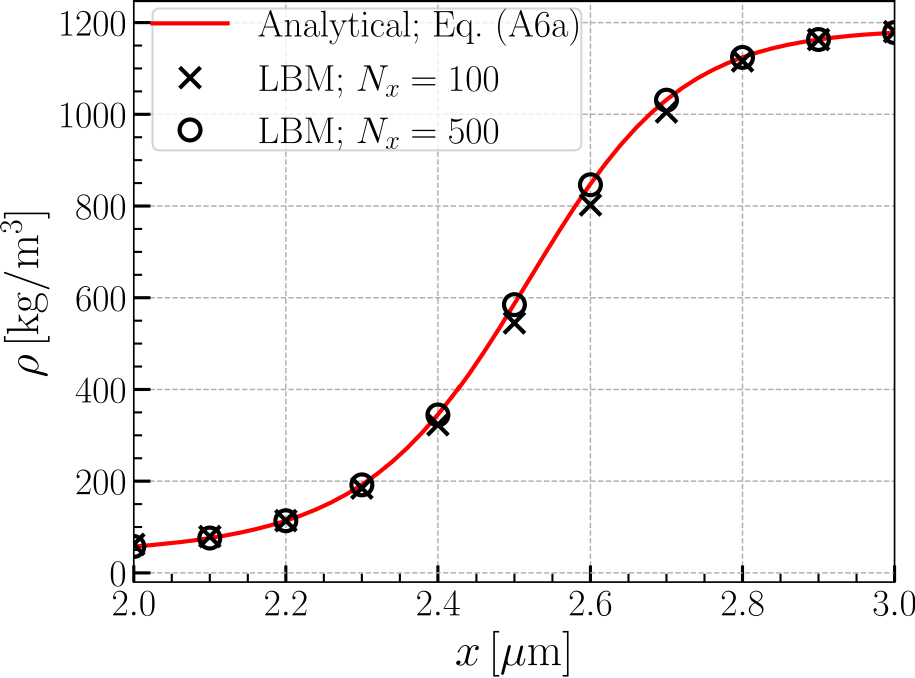}
    \caption{Density profile $\rho$ along the interface region (aligned with the $x$-axis) for the planar interface test. The plot compares analytical and numerical results at different resolutions, denoted by $N_x$.}
\label{fig:Density_Planar}
\end{figure}

Fig.~(\ref{fig:Density_Planar}) shows the density profile along the interface for the coarser mesh ($N_x=100$) and the finer mesh ($N_x=500$). It is observed that for the coarser mesh, there was a small difference compared to the analytical solution. However, the finer mesh coincides with the reference solution with high accuracy.

\subsection{\label{sec:Droplet}Static droplet}

\begin{figure}[ht!]
     \centering
     \includegraphics[width=0.4\textwidth]{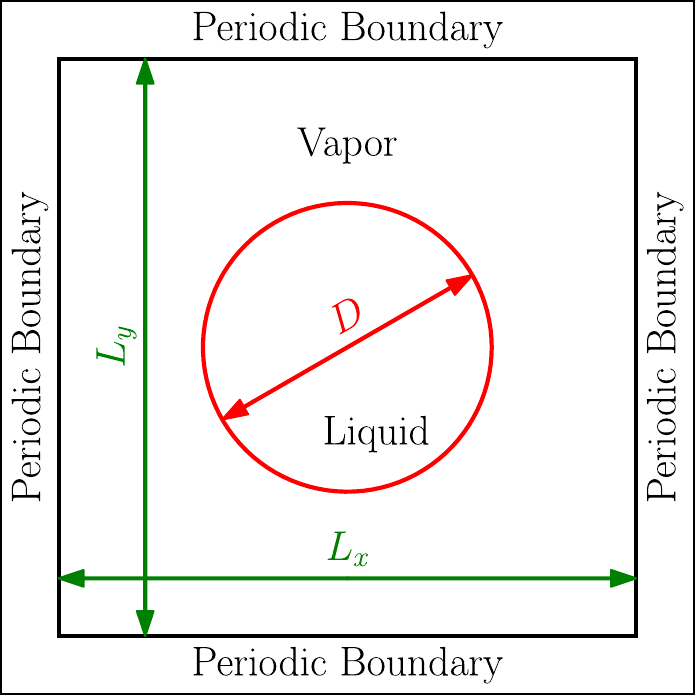}
    \caption{Schematic diagram of the static droplet test configuration. A two-dimensional domain $(L_x \times L_y)$ contains a liquid droplet of diameter $D$ placed at the center, surrounded by vapor. Periodic boundary conditions are applied in all directions.}
\label{fig:DropletSchematic}
\end{figure}

The second test is the static drop test, where a liquid droplet is placed in the middle of a 2D cuboid domain of dimensions $L_x=15~\mathrm{\mu m}$ and $L_y=15~\mathrm{\mu m}$, surrounded by vapor, as shown in Fig.~\ref{fig:DropletSchematic}. Periodic boundary conditions are applied.

First, we assess whether the surface tension reproduced by the method matches the one prescribed by Eq.~\eqref{eq:surface_tension}. To this end, we carry out Young–Laplace test by simulating varying droplet diameters.

\begin{figure}[ht!]
     \centering
     \includegraphics[width=0.48\textwidth]{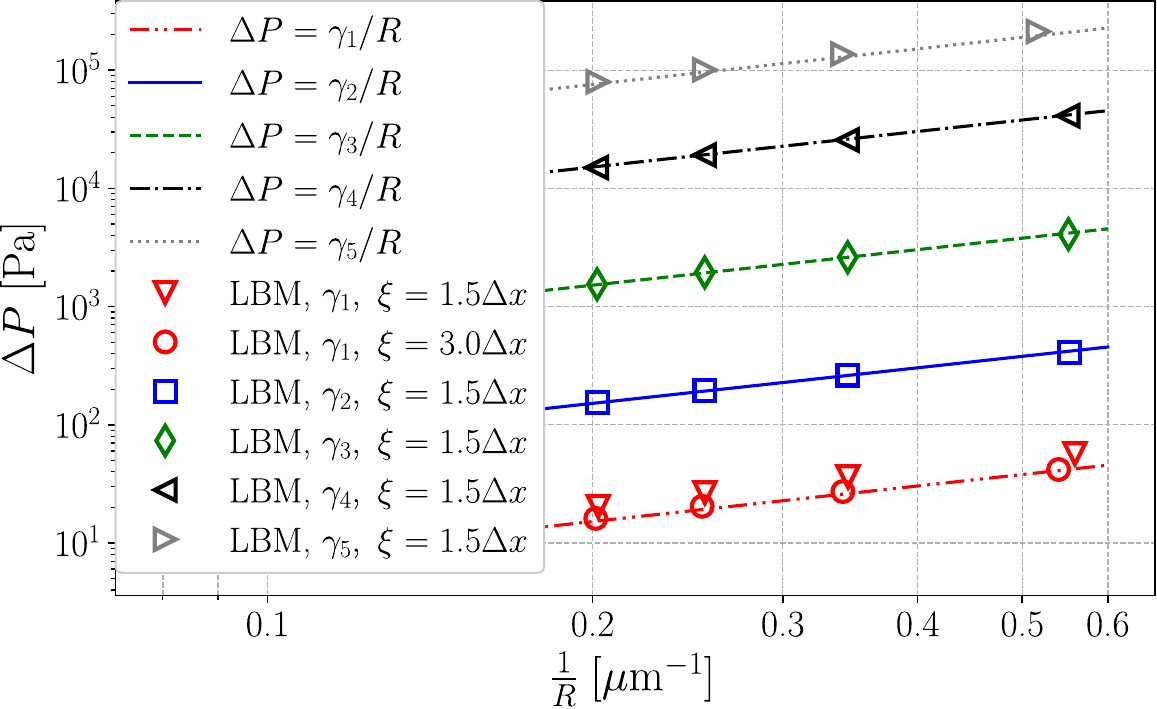}
    \caption{Laplace pressure $\Delta P$ (in Pascal) as a function of droplet curvature $1/R$ for different surface tensions $\gamma_1=0.0758~\mathrm{mN/m}$, $\gamma_2=0.758~\mathrm{mN/m}$, $\gamma_3=7.58~\mathrm{mN/m}$, $\gamma_4=75.8~\mathrm{mN/m}$ and $\gamma_5=379~\mathrm{mN/m}$. Comparison between Young-Laplace relation and LBM results using different interface thickness $\xi$.} 
\label{fig:SurfaceTension}
\end{figure}

When simulating R134a ($\gamma_3=7.58~\mathrm{mN/m}$) using a discretization of $\Delta x = 0.1~\mu\mathrm{m}$ and $\Delta t = 1.26~\mathrm{ns}$, the parameter $\kappa$ obtained from Eq.~\eqref{eq:surface_tension} has a value of 1.026. For values close to one, the term responsible for modifying the surface tension in Eq.~\eqref{eq:interactionForce} tends to zero, indicating that the surface tension is primarily determined by the Shan-Chen force. The results for this case, with $\xi = 1.5\Delta x$, are shown in Fig.~\ref{fig:SurfaceTension}. Even with a small $\xi$, the numerical Laplace pressures are very close to the expected values, indicating that the correct surface tension was imposed.

To test the validity of the proposed approach, we consider surface tensions different from that of R134a. We first test $\gamma_2=0.758~\mathrm{mN/m}$, which is 100 times smaller than the surface tension of water. 
Instead of changing $\kappa$, we employ an alternative strategy to adjust the surface tension: increasing $\tau_{l;\mathrm{lat}}$ from 0.56 to 0.68. Due to the unit conversion in Eq.~\eqref{eq:unit}, this change enables the simulation of the new surface tension while yielding $\kappa = 0.923$, which remains close to one. 
Figure~\ref{fig:SurfaceTension} shows that the results are very good even with $\xi = 1.5\Delta x$. The same approach was applied to simulate $\gamma_4=75.8~\mathrm{mN/m}$, which is close to the surface tension of water. In this case, we decrease $\tau_{l;\mathrm{lat}}$ to 0.52, resulting in $\kappa = 1.140$. The simulation results were also in close agreement with the imposed surface tension.

Finally, we test two additional cases: $\gamma_1=0.0758~\mathrm{mN/m}$ (with $\tau_{l;\mathrm{lat}}=0.68$) and $\gamma_5=379~\mathrm{mN/m}$ (with $\tau_{l;\mathrm{lat}}=0.52$). Instead of further modifying the relaxation time as in the previous cases, we directly adjust $\kappa$ to 0.0923 for the first case and 5.698 for the second. 
As shown in Fig.~\ref{fig:SurfaceTension}, the simulations with $\xi = 1.5\Delta x$ approximate the target surface tensions, but with less accuracy compared to the cases where $\kappa$ is close to one. This suggests that large changes in $\kappa$ introduce numerical errors. 
We repeated the simulation of $\gamma_1$ using a higher resolution with $\Delta x = 0.05~\mu\mathrm{m}$, $\Delta t=0.95~\mathrm{ns}$, and $\xi=3\Delta x$. The finer grid shows better agreement with the desired surface tension in Fig.~\ref{fig:SurfaceTension}, demonstrating the robustness of the proposed approach for imposing surface tension. As a final remark, we attempted to simulate a surface tension of $\gamma_5=758~\mathrm{mN/m}$, but the simulation became unstable due to the high value of $\kappa$.

An important difference between this test and the planar interface test is the presence of curvature in the interface, which leads to spurious currents caused by discretization errors. In this case, we assess the impact of increasing the computational mesh resolution on spurious currents. To this end, we fix $D=5\mathrm{\mu m}$ and $\tau_{l;\mathrm{lat}}=0.62$, and simulate five different meshes, with all relevant data presented in Table~\ref{tab:Droplet_Parameters}.

\begin{table}[ht!]
\centering\small
\renewcommand{\arraystretch}{1.2}
\begin{tabular}{ c | c | c | c | c | c  }
\hline
- & Grid 1 & Grid 2 & Grid 3 &
Grid 4 & Grid 5 \\
\hline
$N_x$ & 100 & 200 & 300 & 
400 & 500 \\
\hline
$N_y$ & 100 & 200 & 300 & 
400 & 500 \\
\hline
$\Delta x$ [nm] & 100 & 50 & 33.3 & 25 & 20 \\
\hline
$\Delta t$ [ns] & 2.532 & 0.633 & 0.281 & 
0.158 & 0.101 \\
\hline
$\xi$ [m] & $1.5\Delta x$ & $3.0\Delta x$ & $4.5\Delta x$ & 
$6.0\Delta x$ & $7.5\Delta x$ \\
\hline
$C_U$ & 39.5 & 79 & 118.5 & 158 & 197.5 \\
\hline
$C_{\gamma}$ & 0.156 & 0.312 & 0.468 & 0.624 & 0.780 \\
\hline

\hline
\end{tabular}
\caption{Discretization parameters of static droplet test case. $N_x$ and $N_y$ are the number of grid nodes used to discretize the domain. $\Delta x$ (in nanometers) is defined as $L_x/N_x$. $\Delta t$ (in nanoseconds) is the time step and $\xi$ is the interface thickness. $C_U$ and $C_{\gamma}$ are the conversion factors from lattice units ($l.u.$) to physical units for velocity and surface tension -- Eq.~(\ref{eq:unit2}). }
\label{tab:Droplet_Parameters}
\end{table}

\begin{figure}[ht!]
     \centering
     \includegraphics[width=0.48\textwidth]{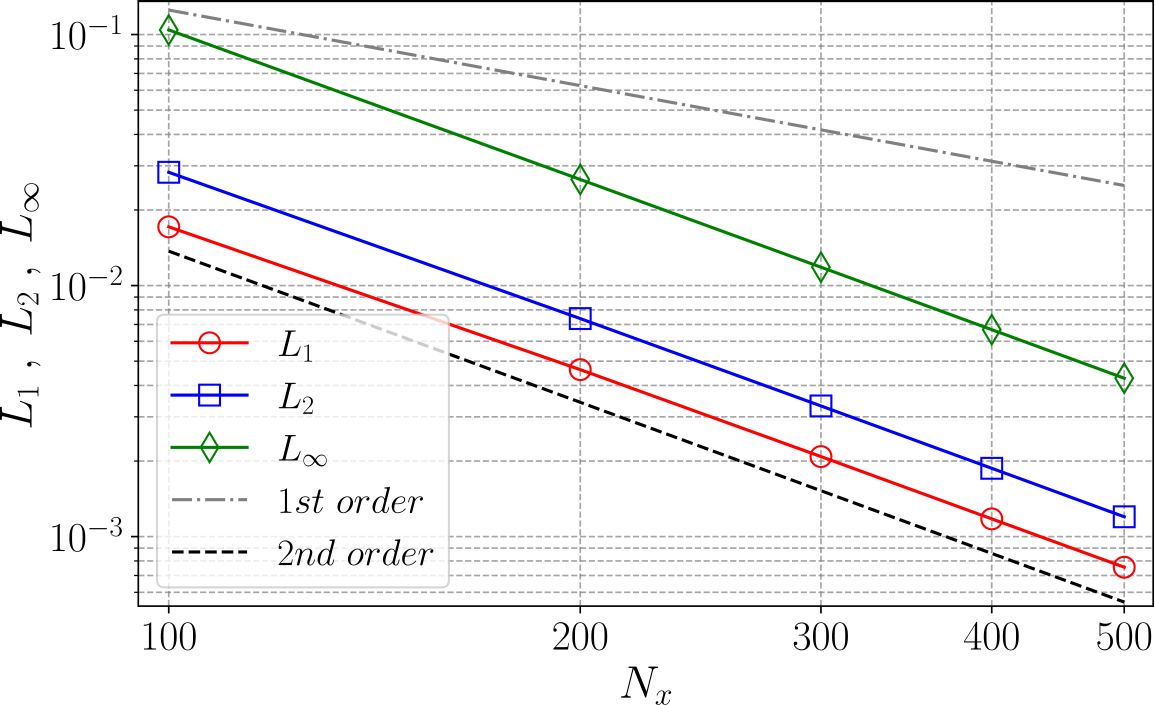}
    \caption{Comparison of the $L_1$, $L_2$, and $L_{\infty}$ norm errors for the static droplet test. The norms, defined in Eq.~(\ref{eq:norms}), are computed from the velocity magnitude field.}
\label{fig:Droplet}
\end{figure}

The simulation results are presented in Fig.~\ref{fig:Droplet}. Norms are computed from the velocity magnitude field using Eq.~(\ref{eq:norms}) since no spurious currents should be observed in reality. The error norms converge with second order of accuracy. 

The simulation with the coarse grid ($N_x=100$), resulted in a maximum spurious currents magnitude of $0.1~\mathrm{m/s}$ and $2.6\times10^{-3}$ lattice unis. With the finest resolution this magnitude decreased to $4.3\times10^{-3}~\mathrm{m/s}$ and $2.2\times10^{-5}$ in lattice units. It is observed that the velocity in physical units converges with second-order accuracy, while the velocity in lattice units converges to zero with third-order accuracy. This occurs because the conversion factor $C_U$ increases linearly with the grid resolution.

It is important to mention that there are numerical schemes designed to eliminate the presence of spurious currents, such as well-balanced scheme\cite{guo2021well} In this work, we employ a discretization that still exhibits spurious currents; however, our intention is to demonstrate the method’s order of convergence through mesh refinement.


\subsection{\label{sec:Galilean}Galilean invariance}

In this section, we describe a test used to evaluate whether the P-LBM respects Galilean invariance. This test is similar to the flat interface test; however, here we introduce a constant velocity $U_x$ throughout the fluid. If the method respects Galilean invariance, the velocity profile should remain constant throughout the simulation. We can consider the fluid to be static relative to an inertial reference frame moving with velocity $U_x$; thus, the results of this simulation should not differ from those of the flat interface test. However, due to discretization errors, the method may exhibit a deviation from this behavior.

\begin{figure}[ht!]
     \centering
     \includegraphics[width=0.48\textwidth]{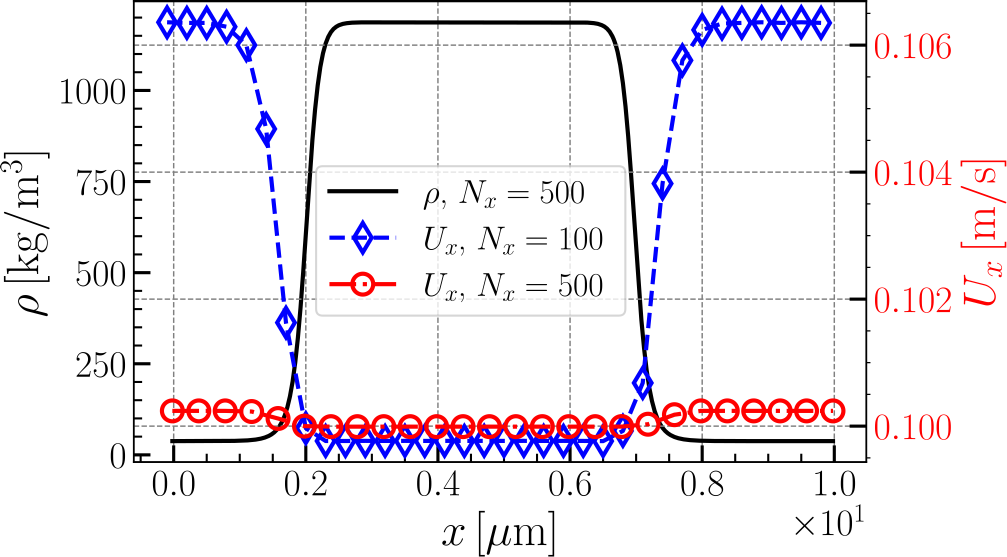}
    \caption{LBM results for Galilean invariance test. Left axis: Density profile $\rho$ along $x$-axis. Right axis: Velocity profile $U_x$ along $x$-axis. Comparison between two grid resolutions $N_x=100$ and $N_x=500$.}
\label{fig:galileanVelocity}
\end{figure}

The configuration of this problem is identical to the flat interface described in Fig.~\ref{fig:PlanarInterfaceSchematic}. Also, simulation parameters are equal to the ones shown in Table~\ref{tab:Planar_Interface_Parameters} with the addition of the velocity $U_x=0.1~\mathrm{m/s}$. The simulation begins with zero velocity until \( t = 7.91~\mathrm{\mu s} \) seconds, allowing the interface density profile to reach equilibrium. A force field is then applied to abruptly change the system's velocity to \( U_x \). The simulation continues until \( t = 31.6~\mathrm{\mu s} \), and the results are recorded.

\begin{figure}[ht!]
     \centering
     \includegraphics[width=0.48\textwidth]{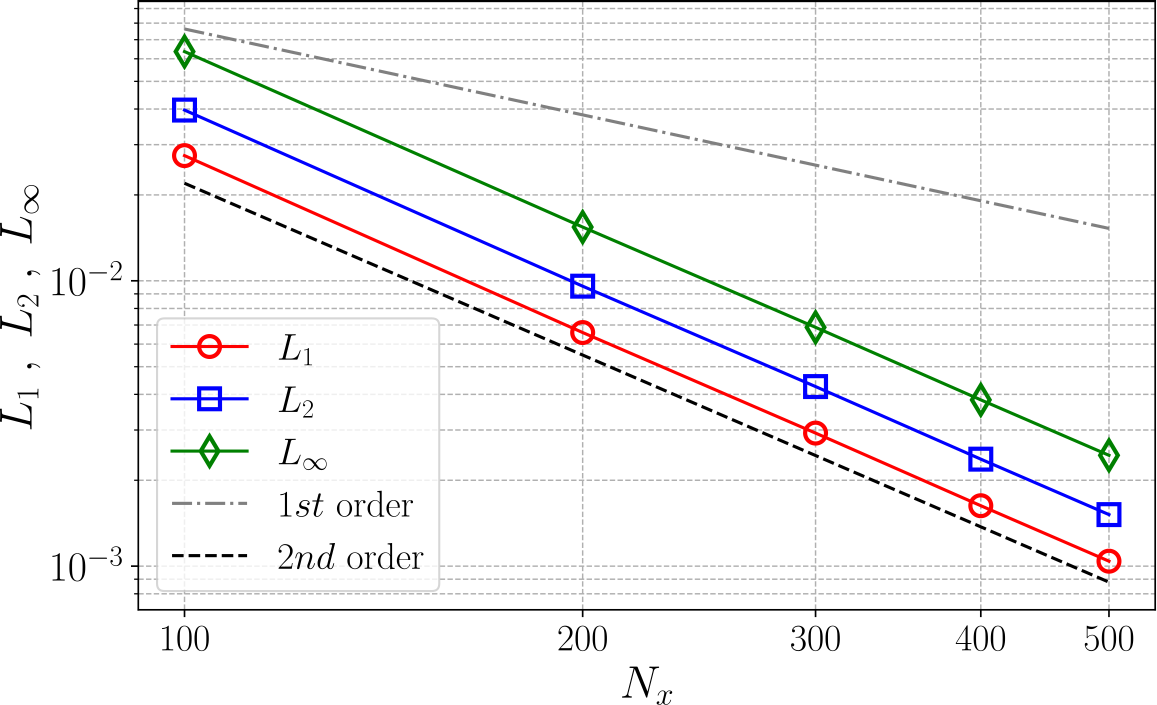}
    \caption{Errors in the $L_1$, $L_2$, and $L_{\infty}$ norms for the Galilean invariance test. The configuration includes a planar interface subjected to a constant velocity field ($U_x = 0.1~\mathrm{m/s}$). }
\label{fig:galilean}
\end{figure}

To illustrate this behavior, we compare the simulation results for $N_x=100$ and $N_x=500$ in Fig.~\ref{fig:galilean}. The fluid velocity is expected to remain constant throughout the domain, but it is observed that the vapor and liquid phases exhibit a relative velocity. This relative motion between phases persists over time, indicating that it is not a transient phenomenon, but rather a consequence of discretization errors acting at the interface. This highlights the importance of mesh convergence analysis to ensure that such effects do not influence the simulation results.

The error norms are presented in Fig.~\ref{fig:galilean}.
The norms are given by Eq.~(\ref{eq:relative_norms}) and are computed from the velocity field. The analytical solution is simply a constant field with magnitude $U_x$. The error norms converge with second order of accuracy. 

\subsection{\label{sec:Flow}Two phase flow between parallel plates}

\begin{figure}[ht!]
     \centering
     \includegraphics[width=0.4\textwidth]{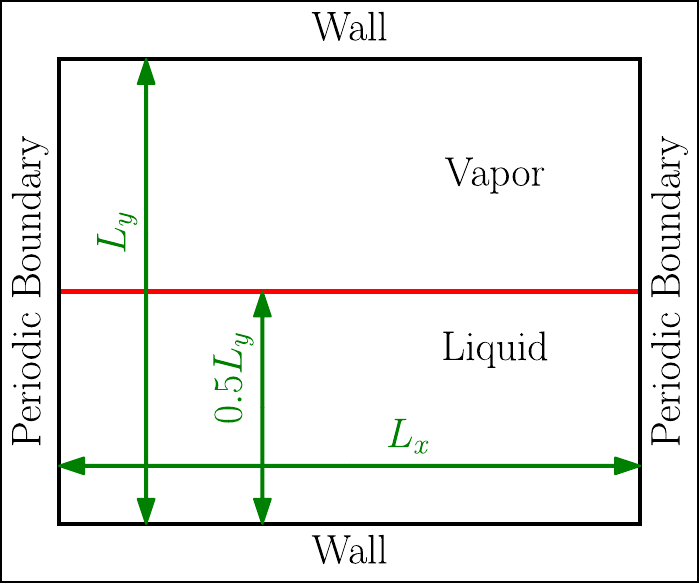}
    \caption{Schematic diagram of the two-phase flow test between parallel plates. A two-dimensional domain $(L_x \times L_y)$ features solid walls at the top and bottom boundaries and periodic boundary conditions on the sides. The lower part contains liquid, while the upper part contains vapor.}
\label{fig:ParallelFlow}
\end{figure}

The last test is the two-phase flow between parallel plates. A schematic describing the geometry of the simulation is shown in Fig.~\ref{fig:ParallelFlow}. Again, a 2D cuboid geometry with dimensions $L_x=0.5~\mathrm{\mu m}$ and $L_y=10~\mathrm{\mu m}$ is defined. Periodic boundary conditions are applied to the side boundaries, while solid walls are applied to the lower and upper boundaries. The bottom part of the domain is filled with liquid, while the upper part is filled with vapor. 
A volumetric force $F_x=10^6~\mathrm{N/m^3}$ is applied in the entire domain to drive the flow.

Since periodic boundary conditions are applied to the side boundaries, it is not necessary to have a long domain length. Therefore, we choose \( L_x < L_y \) to save computational time.
Five grid resolutions are tested and parameters are provided in Table~\ref{tab:Multiphase_Table}.

\begin{table}[ht!]
\centering\small
\renewcommand{\arraystretch}{1.2}
\begin{tabular}{ c | c | c | c | c | c }
\hline
- & Grid 1 & Grid 2 & Grid 3 & Grid 4 & Grid 5 \\
\hline
$N_x$ & 5 & 10 & 15 & 20 & 25 \\
\hline
$N_y$ & 100 & 200 & 300 & 
400 & 500 \\
\hline
$\Delta x$ [nm] & 100 & 50 & 33.3 & 25 & 20 \\
\hline
$\Delta t$ [ps] & 
1266 & 316.4 & 
140.7 & 79.11 & 50.63 \\
\hline
$\xi$ [$m$] & $1.5\Delta x$ & $3.0\Delta x$ & 
$4.5\Delta x$ & $6.0\Delta x$ & $7.5\Delta x$ \\
\hline
$C_U$ [m/s per l.u.] & 79 & 158 & 237 & 316 & 395 \\
\hline
$C_{VF}$ [$\mathrm{TN/m^3}$ per l.u.] & 62.41 & 499.3 & 1685 & 3994 & 7801 \\
\hline
$C_{\gamma}$ [N/m per l.u.] & 0.62 & 1.25 & 1.87 & 2.50 & 3.12 \\
\hline
\end{tabular}
\caption{Discretization parameters of two phase flow between parallel plates test.
$N_x$ and $N_y$ are the number of grid nodes used to discretize the domain. $\Delta x$ (in nanometers) is defined as $L_x/N_x$. $\Delta t$ (in picoseconds) is the time step and $\xi$ is the interface thickness. $C_U$, $C_{VF}$ and $C_{\gamma}$ are the conversion factors from lattice units ($l.u.$) to physical units for velocity, volumetric force and surface tension -- Eq.~(\ref{eq:unit2}).}
\label{tab:Multiphase_Table}
\end{table}

\begin{figure}[ht!]
     \centering
     \includegraphics[width=0.48\textwidth]{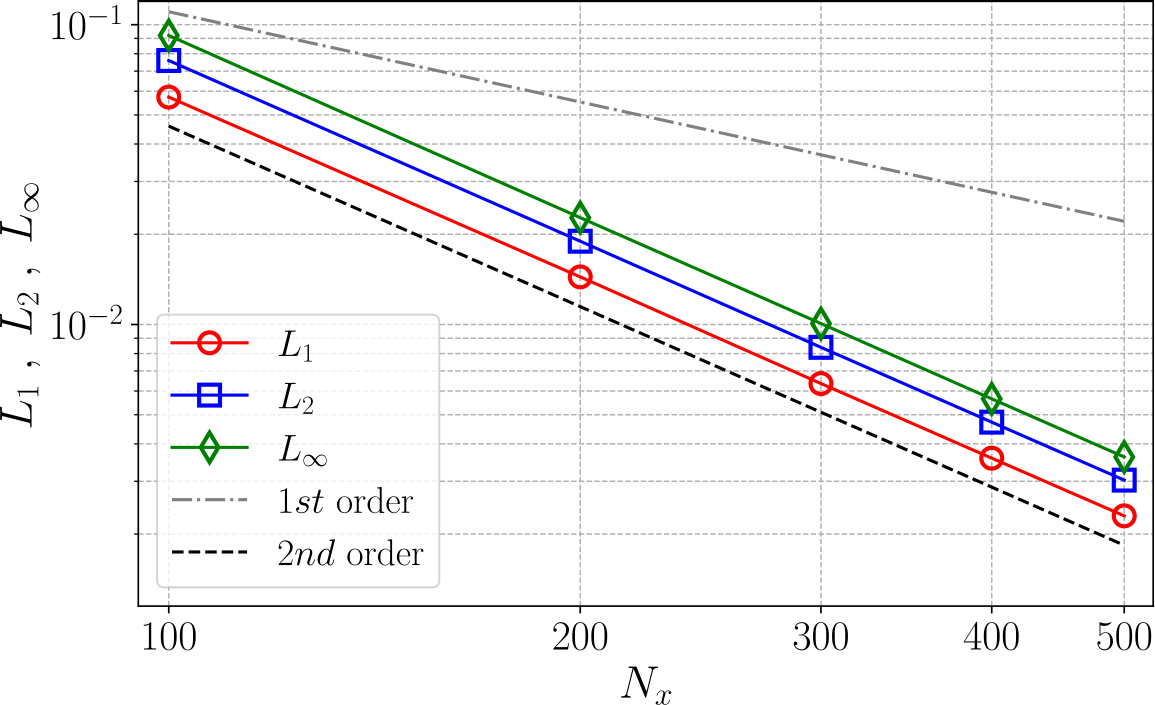}
    \caption{Errors in the $L_1$, $L_2$, and $L_{\infty}$ norms for the two-phase flow test between parallel plates. The norms were computed from Eq.~(\ref{eq:relative_norms}) using the velocity field.}
\label{fig:Multiphase_Flow}
\end{figure}

The simulation results are presented in Fig.~\ref{fig:Multiphase_Flow}.
The norms are given by Eq.~(\ref{eq:relative_norms}) and are computed from the velocity field. Similarly to the previous tests, the error norms converge with second order of accuracy. 

\begin{figure}[ht!]
     \centering
     \includegraphics[width=0.48\textwidth]{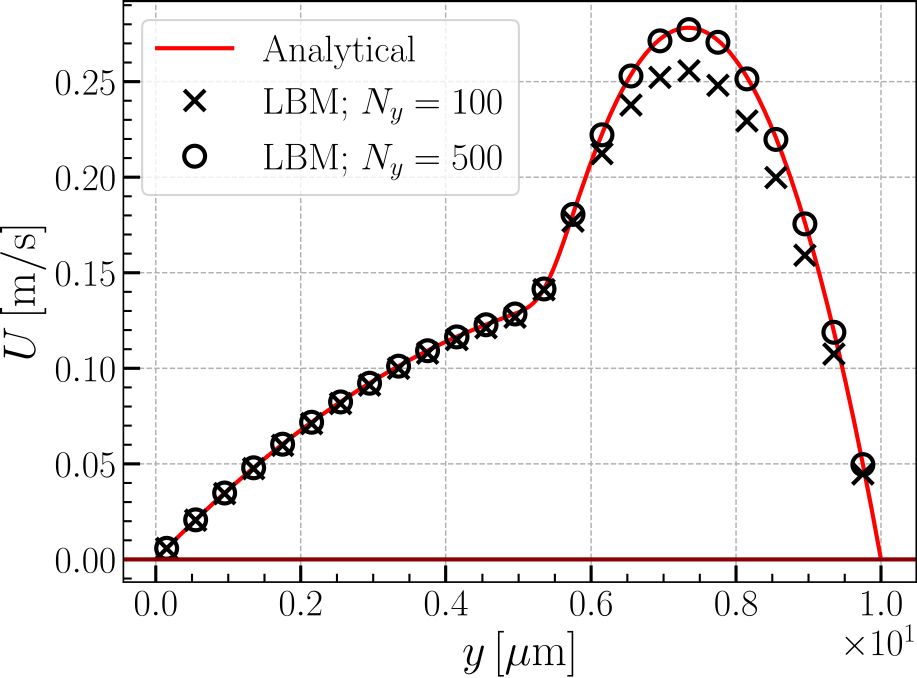}
    \caption{Velocity profile $U$ along the channel height (aligned with the $y$-axis) for the two-phase flow test. The plot compares analytical and numerical results at different resolutions, denoted by $N_y$.}
\label{fig:Multiphase_Plot}
\end{figure}

To illustrate the simulation results, we plot both the numerical solution (with two grid resolutions) and the analytical solution in Fig.~\ref{fig:Multiphase_Plot}. As shown, the highest resolution closely matches the analytical results.

\section{\label{sec:Conclusion}Conclusion}

In this study, we analyzed the consistency of the P-LBM, focusing on its relationship with macroscopic properties and the PDE it approximates. We derived a formulation that directly links the method parameters to physical properties such as phase densities, interface thickness, and surface tension. This approach eliminates the need for empirical parameter tuning, thereby simplifying the application of the method and improving its robustness.

Furthermore, we implemented and validated our approach in OpenLB, introducing a dedicated unit converter for multiphase problems. The benchmarks conducted—including the planar interface, static droplets, Galilean invariance, and two-phase flow between parallel plates—demonstrated second-order convergence and confirmed the numerical consistency of the method. All tests were performed using R134a as working fluid with all properties given in SI units.

The tests performed show that numerical errors occur in multiphase simulations with P-LBM. Therefore, it is important to perform a mesh study to ensure that the numerical errors converge to an acceptable value and to guarantee the reliability of the simulation.

The results obtained not only validate the proposed methodology but also emphasize its potential for future applications in complex and realistic multiphase simulations. The methodology offers a practical approach to applying the pseudo-potential, enabling users to utilize the method more effectively with real simulation parameters.


\section*{Acknowledgments}

We acknowledge the support of the Alexander von Humboldt Foundation in sponsoring the postdoctoral fellowship of researcher Dr.\ Luiz Eduardo Czelusniak at LBRG, KIT.

\section*{Authors Declarations}

\subsection*{Conflict of Interest}
The authors have no conflicts to disclose.

\subsection*{Data Availability Statement}
The data that support the findings of this study are available from the corresponding author upon reasonable request. In addition, the codes used in this work will be available in the next release of OpenLB (1.8), which can be downloaded from the website: \url{https://www.openlb.net/download/}.

\appendix 

\section{\label{sec:Interface}Interface profile}

In the following analysis, two fluids separated by a flat interface at equilibrium are considered. Density gradients exist only along the x-axis. The component $p_{xx}$ of the pressure tensor, Eq.~(\ref{eq:New_Pressure_Tensor}), can be written as:
\begin{equation} \label{eq:xx_simplified}
p_{xx} = p_{EOS} + \frac{1}{6r}
\left[
\left( \frac{d\sqrt{\rho}}{dx} \right)^2
- \sqrt{\rho}\frac{d^2\sqrt{\rho}}{dx^2}
\right].
\end{equation}
Here we consider $\epsilon=2$ as discussed in Section~\ref{sec:Parameters}.

To further simplify the equation, we define $z=d\sqrt{\rho}/dx$.
Next, we assume
that $z$ is a function of $\rho$ to apply the following transformation:
\begin{equation}
\rho^2\frac{d}{d\rho}
\left( \frac{z^2}{\rho} \right) = - \left( \frac{d\sqrt{\rho}}{dx} \right)^2
+ \sqrt{\rho}\frac{d^2\sqrt{\rho}}{dx^2}.
\end{equation}
Then, we arrive at the following ordinary differential equation:
\begin{equation} \label{eq:EDO}
\frac{d}{d\rho}
\left( \frac{z^2}{\rho} \right) = 6r\frac{p_{EOS}-p_{xx}}{\rho^2}.
\end{equation}
From Eq.~(\ref{eq:MomentumConservation}) we notice $p_{xx}=\text{constant}$ for a static planar interface in equilibrium. Also, we have the boundary conditions $p_{xx}=p_{EOS}(\rho_v)=p_{EOS}(\rho_l)$. 

Eq.~(\ref{eq:EDO}) - together with the EOS of Eq.~(\ref{eq:EOS}) and the assigned boundary conditions have the following solution:
\begin{equation} \label{eq:Interface_Derivative}
z = - \sqrt{\frac{6rp_c}{\rho_c^3}} 
(\sqrt{\rho}-\sqrt{\rho_v})(\rho-\rho_l),
\end{equation}
with $z=d\sqrt{\rho}/dx$. This equation can also be written as:
\begin{equation}
\sqrt{\frac{6rp_c}{\rho_c^3}} dx = - \frac{ d\sqrt{\rho} }
{ (\sqrt{\rho}-\sqrt{\rho_v})(\rho-\rho_l) }.
\end{equation}
Integrating this equation, we obtain:
\begin{subequations} \label{eq:analytical_interface}
\begin{equation}
\sqrt{\frac{6rp_c}{\rho_c^3}} x + C_0 = f_{\xi}(\rho),
\end{equation}
where $C_0$ is an integration constant and: 
\begin{equation}
\begin{aligned}
f_{\xi}(\rho) &= \frac{1}{\rho_l-\rho_v}\text{ln}
\left( \sqrt{\rho}-\sqrt{\rho_v} \right), \\
&- \frac{1}{2\sqrt{\rho_l}(\sqrt{\rho_l}-\sqrt{\rho_v})}\text{ln}
\left( \sqrt{\rho_l}-\sqrt{\rho} \right), \\
&- \frac{1}{2\sqrt{\rho_l}(\sqrt{\rho_v}+\sqrt{\rho_l})}\text{ln}
\left( \sqrt{\rho} + \sqrt{\rho_l} \right).
\end{aligned}
\end{equation}
\end{subequations}

We define the interface width $\xi$ as the distance where $\rho_v+0.33(\rho_l-\rho_v)<\rho<\rho_v+0.67(\rho_l-\rho_v)$.
Then: 
\begin{equation}
\xi = \sqrt{\frac{\rho_c^3}{6rp_c}}
(f_{\xi}(\rho_{I;1})-f_{\xi}(\rho_{I;2})),
\end{equation}
where $\rho_{I;1}=\rho_v+0.33(\rho_l-\rho_v)$ and $\rho_{I;2}=\rho_v+0.67(\rho_l-\rho_v)$.


\section{\label{sec:Surface}Surface tension}

The surface tension in diffuse interface models for a planar interface case is defined as\cite{rowlinson1982molecular}:
\begin{equation}
\gamma = \int_{-\infty}^{+\infty} (p_{xx}-p_{yy}) dx,
\end{equation}
where $x$ and $y$ represents the normal and tangential directions in respect to the interface. Considering the pressure tensor expression - Eq.~(\ref{eq:New_Pressure_Tensor}) - we have the formula:
\begin{equation}
\gamma = - \frac{\kappa}{9r} \int_{-\infty}^{+\infty} \sqrt{\rho} \frac{d^2\sqrt{\rho}}{dx^2}  dx.
\end{equation}

Using integral by part, we rewrite the above expression:
\begin{equation}
\gamma = \frac{\kappa}{9r} \left( - \left. \sqrt{\rho} \frac{d\sqrt{\rho}}{dx} \right|_{-\infty}^{+\infty} + \int_{-\infty}^{+\infty} \left( \frac{d\sqrt{\rho}}{dx} \right)^2   dx
\right).
\end{equation}

The interface density is constant at the bulk phases meaning $d\sqrt{\rho}/dx=0$ for $x=\pm\infty$. Then, we write the final form for the expression:
\begin{equation}
\gamma = \frac{\kappa}{9r} \int_{\rho_v}^{\rho_l}  \frac{d\sqrt{\rho}}{dx} d\sqrt{\rho}.
\end{equation}
We already know the expression for $d\sqrt{\rho}/dx$ which is equal to $z(\rho)$ in Eq.~(\ref{eq:Interface_Derivative}). Then, the 
surface tension is simply obtained by integration:
\begin{equation}
\gamma = \sqrt{\frac{2rp_c}{27\rho_c^3}} \kappa \left( f_{\gamma}(\rho_v)-f_{\gamma}(\rho_l) \right).
\end{equation}
with:
\begin{equation}
f_{\gamma}(\rho) = \frac{\rho^2}{4} - \frac{\rho_l}{2}\rho 
- \frac{\sqrt{\rho_v}}{3}\rho^{1.5} + \sqrt{\rho_v}\rho_l \rho^{0.5}.
\end{equation}



\providecommand{\noopsort}[1]{}\providecommand{\singleletter}[1]{#1}%

\end{document}